\documentstyle{mn}
\input{epsf}

\title[Colour-colour diagrams of LMXBs]
{A comment on the colour-colour diagrams of Low Mass X-ray Binaries}

\author[M. Gierli\'nski, C. Done] 
{Marek~Gierli\'nski$^{1,2}$ and Chris Done$^1$\\
$^1$Department of Physics, University of Durham, South Road, Durham DH1 3LE, 
UK\\ 
$^2$Obserwatorium Astronomiczne Uniwersytetu Jagiello{\'n}skiego, 30-244 
Krak{\'o}w, Orla 171, Poland}

\date{Submitted to MNRAS}
\pagerange{\pageref{firstpage}--\pageref{lastpage}}
\pubyear{2001}

\begin{document}

\topmargin = -0.5cm

\maketitle

\label{firstpage}

\begin{abstract}

Disc accreting neutron stars come in two distinct varieties, atolls 
and Z sources, named after their differently shaped tracks on a 
colour-colour diagram as the source luminosity changes. Here we 
present analysis of three transient atoll sources showing that there 
is an additional branch in the colour-colour diagram of atoll sources 
which appears at very low luminosities. This new branch connects to 
the top of previously known C-shaped (atoll) path, forming a 
horizontal track where the average source flux decrease from right to 
left. This turns the C-shape into a Z. Thus both atolls and Z sources 
share the same topology on the colour-colour diagram, and evolve in 
similar way as a function of increasing averaged mass accretion rate. 
This strongly favours models in which the underlying geometry of 
these sources changes in similar ways. A possible scenario is one 
where the truncated disc approaches the neutron star when the 
accretion rate increases, but that in the atolls the disc is 
truncated by evaporation (similarly to black holes) whereas in the Z 
sources it is truncated by the magnetic field.
  
\end{abstract}


\section{Introduction}
\label{sec:introduction}

Low mass X-ray binaries (LMXBs) hosting a neutron star can be 
observationally divided into two main categories, dubbed atolls and Z 
sources (Hasinger \& van der Klis 1989). This classification is based 
on changes in both spectral and timing properties as the source 
varies. Z sources are named after a Z-shaped track they produce on an 
X-ray colour-colour diagram. Atolls can fall into one of the two 
spectral states, a hard, low luminosity `island' or a soft, high 
luminosity `banana'. They trace a U-shaped or C-shaped track as the 
source spectrum evolves between the island and the banana  (see e.g.\ 
fig.\ 1 in M{\'e}ndez et al.\ 1999). These differences between the 
two LMXBs categories probably reflect differences in both mass 
accretion rate, $\dot{M}$, and magnetic field, $B$, with the Z 
sources having high luminosity (typically more than 50 per cent of 
the Eddington limit) and magnetic field ($B \ge 10^9$ G) while the 
atolls have lower luminosity (generally less than 10 per cent of 
Eddington) and low magnetic field ($B \ll 10^9$ G) (Hasinger \& van 
der Klis 1989). 

Both atolls and Z sources move {\em along\/} their tracks on the 
colour-colour diagram and do not jump between the track branches. 
Most of the X-ray spectral and timing properties, e.g.\ the kilohertz 
quasi-periodic oscillation (QPO) frequency (M{\'e}ndez \& van der 
Klis 1999), depend only on the position of a source in this diagram. 
This is usually parameterized by the curve length, $S$, along the 
track. 

\begin{figure*}
\begin{center} 
\leavevmode 
\epsfxsize=16cm \epsfbox{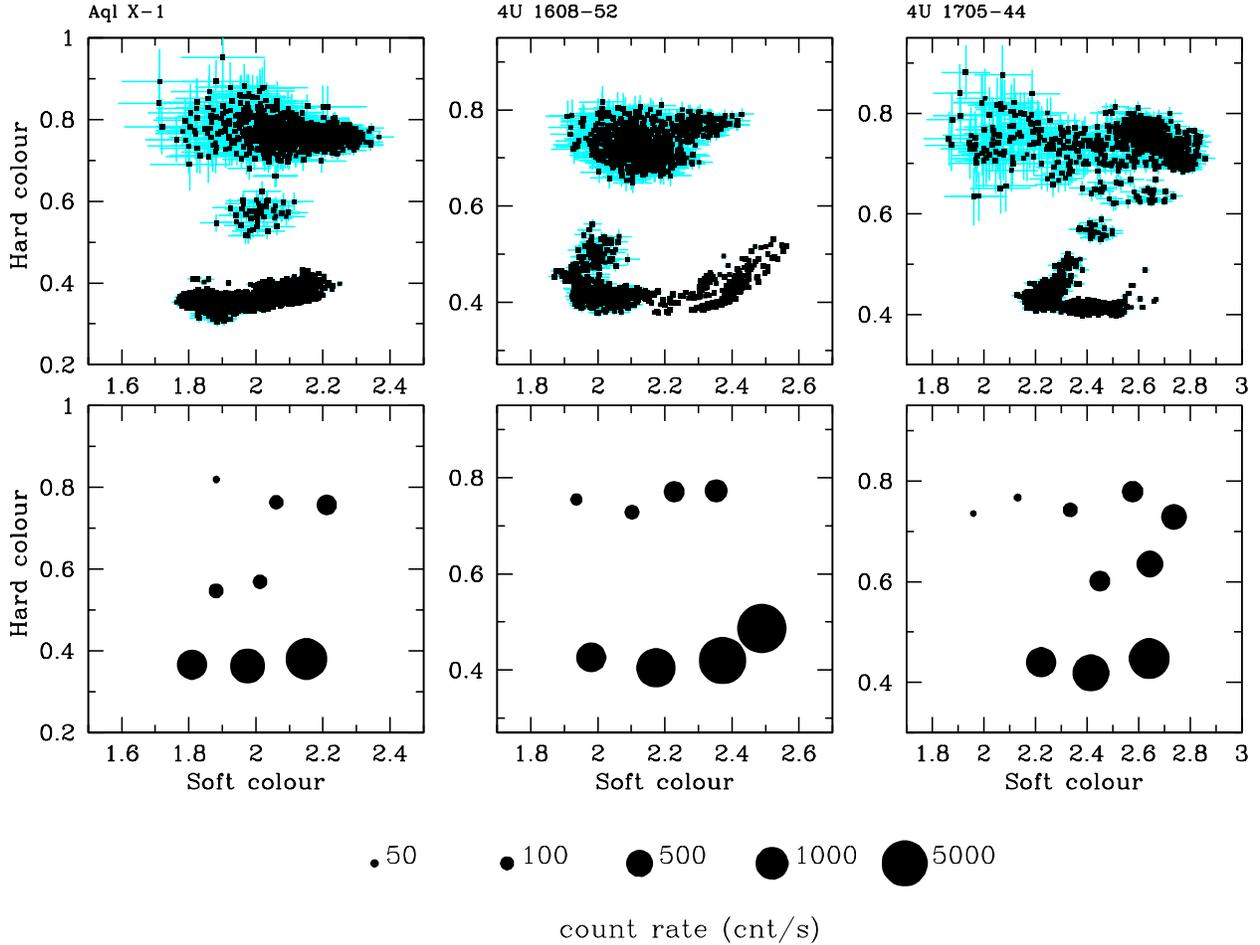} 
\end{center} 
\caption{Colour-colour diagrams of three atoll sources. The colours 
are defined as a ratios of the counts in 4--6.4 keV over 3--4 keV 
bands (soft) and 9.7--16 over 6.4--9.7 keV bands (hard). The upper 
panels show original background-subtracted diagrams with the bin 
size of 128 s. The lower panels contain rebinned data and a size of 
each data point is proportional to the total count rate in detectors 
0, 2 and 3, in the 3--16 keV band.} 
\label{fig:colcol} 
\end{figure*}

This strongly suggests that a single parameter determines the overall 
properties of the source, generally believed to be the accretion 
rate, which increases with $S$ from the horizontal (top) to the 
flaring (bottom) branch in the Z sources, and from the island to the 
banana in the atolls. The increasing characteristic frequencies in 
the power spectra along the track are then generally explained by the 
inner disc radius decreasing as a function of $S$ (e.g.\ the review 
by van der Klis 2000).  However, the situation is more complex as $S$ 
is not simply related to the observed X-ray luminosity as would be 
expected if it is determined by the mass accretion rate (e.g. van der 
Klis 2000; 2001).

In this letter we present a compilation of {\it RXTE\/} data from 
three transient atoll sources. Their large amplitude of luminosity 
variation allows us to plot the full track in the colour-colour 
diagram, which appears to form a shape of `Z', similar to Z sources.

\section{The data}
\label{sec:data}

We have analysed the {\it RXTE\/} observations of the three atoll 
LMXBs: Aql X-1 (Cui et al.\ 1998; Reig et al.\ 2000), 4U~1608--52 
and 4U~1705--44 (Hasinger \& van der Klis 1989). 4U~1608--52 is a 
transient source and its most recent outburst in 1998 was observed 
by {\it RXTE\/} with a good coverage of both island and banana 
states. Aql X-1 is also a transient, showing outbursts in a 
time-scale of months to years. 4U~1705--44 is a strongly variable 
X-ray 
source, switching between the island and 
banana states on the time-scales of months.

We have used publicly available {\it RXTE}/PCA data of these three 
sources from PCA epochs 3 (between 1996 April 15 and 1999 March 22) 
and 4 (between 1999 March 22 and 2000 May 13). We selected data from 
detectors 0, 2 and 3, excluding all type I bursts and observations 
with very poor statistics. This gave 477 ks of data for Aql X-1, 219 
ks for 4U~1608--52 and 172 ks for 4U~1705--44. The PCA light-curves 
for each energy channel were extracted in 128 s bins. These were 
used to build a colour-colour diagram, defining a soft colour as a 
ratio of 4--6.4 to 3--4 keV count rates, and a hard colour as a 
9.7--16 over 6.4--9.7 keV ratio. 

The response of the PCA detectors is slowly varying as a function of 
time. Additionally, the high-voltage settings of the instruments were 
altered between the PCA epochs 3 and 4. Therefore, the energy 
boundaries of each energy channel change in time, causing a shift in 
the colours. We have approximately taken these changes into account 
by reading these boundaries from the response matrices created for 
the beginning and end of each PCA epoch, and by linearly 
interpolating between them. When accumulating counts in each of the 
four energy bands (used for computing the colours), we have 
interpolated the number of counts for channels on the edges of these 
energy bands. We have checked the correctness of this procedure using 
Crab data from various observations in both PCA epochs. We computed 
colours and noticed that the position of Crab on the diagram still 
changes in time. This is due to the approximate character of the 
colours we had calculated. Therefore, to account for this variation, 
we have calculated multiplicative factors for the colours from the 
Crab data and applied them to our data (see e.g.\ Homan et al.\ 2001 
for the similar method). The final result is presented in Fig.\ 
\ref{fig:colcol} (upper panels).

To make these plots clearer and to enhance the evolutionary track on 
the diagram we have rebinned the data using a nearest neighbour 
clustering technique. In each step of the iteration the two nearest 
bins (data points) on the colour-colour diagram were found. The total 
numbers of counts in both bins were added and a new count rate and 
colours were calculated. Thus, the two bins were replaced by one. The 
procedure was carried on until the assumed number of bins (between 8 
and 10, as seen in the lower panels of Fig.\ \ref{fig:colcol}) was 
reached. This method gathers together the data with similar spectral 
properties (colours) as opposed to increasing the length of the time 
bin, which averages data in time. The result is presented in Fig.\ 
\ref{fig:colcol} (lower panels). The size of the symbol is 
proportional to the logarithm of the count rate, so it gives an 
overview of the luminosity changes in the diagram.

\section{Results}
\label{sec:results}

All three colour-colour diagrams in Fig.\ \ref{fig:colcol} show 
features characteristic of atoll sources that have been known for 
many years (e.g.\ van der Klis 1995; M{\'e}ndez 1999). There is a 
banana in the lower part and several islands in the upper part of the 
diagram. The banana is significantly brighter than the islands, and 
the X-ray flux increases along the banana branch, from left to right.

However, this large compilation of data shows other features which 
have not previously been seen from sparser data sets. There are {\em 
three\/} distinct branches in the colour-colour diagram.  The banana 
branch forms a lower horizontal track at hard colour of $\sim$ 0.4, 
while the island state begins along the diagonal track which connects 
to the left-hand end of the banana. But there is a part of the 
diagram not reported before: an upper horizontal branch, at hard 
colour of $\sim$ 0.7--0.8, connected to the upper right end of the 
diagonal branch. It is particularly pronounced in 4U~1705--44. This 
extends the previously known C-shaped pattern into a Z. There were 
hints of this extension in previous observations of 4U~1705--44 
(Langmeier et al.\ 1989), but this is the first time that it is shown 
so noticeably.

\begin{figure}
\begin{center} 
\leavevmode 
\epsfxsize=6cm \epsfbox{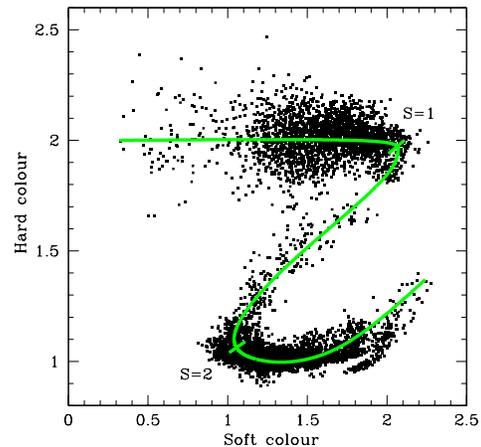} 
\end{center} 
\caption{The `big picture' of the atoll sources: a combined 
colour-colour diagram of Aql X-1, 4U~1608--52 and 4U~1705--44. The 
colours of the three sources from Fig.\ \ref{fig:colcol} were 
linearly transformed so as to co-align all three systems at turns on 
the track. The sources clearly define a Z-shaped track. The curve 
defines the position along the track, $S$.} 
\label{fig:bigcol} 
\end{figure}

To show this new branch more clearly we have created a combined 
colour-colour diagram of all three sources. Since these three atolls 
have similar spectral and timing properties, they probably have the 
same underlying accretion geometry and radiation mechanisms. Thus we 
might expect their spectral evolution, and hence the colour-colour 
diagrams to be the same, but despite similarities in shape, the 
colour-colour diagrams of Fig.\ \ref{fig:colcol} are {\em shifted\/} 
with respect to each other. This is mainly caused by differences in 
absorption, which strongly affects the soft colour, but there might 
also be subtle shifts in hard colour from differences in spin 
frequency (affecting disc-to-boundary layer luminosity ratio and 
hence the soft colour as well) and perhaps also the inclination 
angle.  We have attempted to correct for these effects by simply 
shifting the diagrams in both colours. We have linearly transformed 
each 128 s data point of each source in a way that the left-hand edge 
of the lower branch have transformed colours of (1, 1) and the 
right-hand edge of the upper branch transforms to (2, 2). The 
resulting diagram of all three sources together is shown in Fig.\ 
\ref{fig:bigcol}. We can see that the three diagrams coincide very 
well, and the three branches forming a Z-shaped track are clearly 
visible.

We have analysed source movement along the Z-shaped track. For this, 
we have taken the original (i.e.\ not colour-binned) colour-colour 
diagram of 4U~1705--44 and traced how the position of the source (in
128~s data bins) moves with time. We confirm that motion in the 
diagram, including the upper branch, goes {\em along\/} the track and 
we do not notice jumps between the branches (though jumps cannot be 
completely excluded, since the data is rather sparse in time). 4U 
1705--44 can cross the full width of the upper branch in about 10 
days, while transition on the diagonal takes about five days (a 
similar transition in 4U~1608--52 took about three days).  The 
movement in the lower branch is much faster, with time-scales of 
hours.

\begin{figure}
\begin{center} 
\leavevmode 
\epsfxsize=6cm \epsfbox{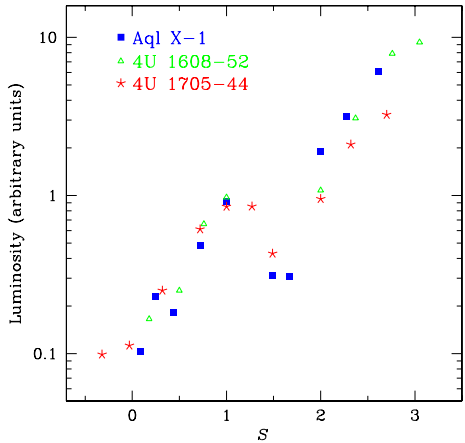} 
\end{center} 
\caption{Bolometric luminosity (absorption corrected) as a function 
of $S$, i.e. distance along the Z track. The luminosities are 
linearly scaled so that they match at the $S=1$ point. The evolution 
of luminosity as a function of $S$ is remarkably similar for all 3 
sources.}
\label{fig:flux} 
\end{figure}

The lower panels in Fig.\ \ref{fig:colcol} show that the average 
X-ray flux at a given point on the track increases from left to right 
(i.e.\ with increasing soft colour) both in the upper and lower 
branches. To study the source flux along the track in details, for 
each grouped point in Fig.\ \ref{fig:colcol} we have extracted a 
typical PCA spectrum from a single {\it RXTE\/} pointing (one to 
three orbits, usually 3--10 ks of exposure) with the same colours and 
mean count rate. These spectra were fit with a model of a blackbody, 
Comptonized component and its reflection, with absorption set at 0.5, 
1.5 and 1.2$\times 10^{22}$ cm$^{-2}$, for Aql X-1, 4U~1608--52 and 
4U~1705--44 respectively (Church \& Ba{\l}uci{\'n}ska-Church 2001; 
Penninx et al.\ 1989; Predehl \& Schmitt 1995). These show that the 
spectra all along the upper horizontal branch are rather hard, 
similar to those previously seen in the hardest island states. The 
bolometric flux was calculated by extrapolating the unabsorbed model 
spectrum, so it is somewhat model dependent but gives at least a 
zeroth order correction. In Fig.\ \ref{fig:flux} we plot this 
bolometric flux in arbitrary units as a function of distance $S$ 
along the Z track. We have defined the right-hand edge of the upper 
branch as $S=1$, and the left-hand edge of the lower branch as $S=2$ 
and linearly interpolated to get the $S$ value of all the other data 
points (see Fig.\ \ref{fig:bigcol}). The bolometric fluxes are scaled 
so that they are roughly equal at the $S=1$ point. It is clear that 
the shapes of these curves are very similar for all 3 sources, 
showing a steady rise in the upper branch ($S \le 1$), and then a 
drop on the diagonal branch, and then rising again along the lower 
branch.

This {\em average\/} bolometric flux most likely corresponds to the 
average mass accretion rate. Then, extending a LMXB paradigm we 
suggest that the accretion rate increases along the Z-shaped track in 
the colour-colour diagram. There is however a {\it decrease\/} of the 
flux on the diagonal branch. We speculate that this might associated 
with jet formation. It is known that jets are associated with state 
transitions in both neutron star and black hole transients (e.g. 
Fender \& Kuulkers 2001). Alternatively, the mass accretion rate onto 
the central source (and hence the hard X-ray flux) could be reduced 
if much of the inflowing material was used to extend the disc 
inwards.

To convert X-ray fluxes into true luminosity requires a distance. For 
these three sources the best distance estimates are: 3.6~kpc for 4U 
1608--52 from radius expansion of X-ray bursts (Nakamura et al.\ 
1989), 4--6.5~kpc for Aql X-1 from optical spectroscopy of the 
companion star (Rutledge et al.\ 2001) and 6.3--8.2~kpc for 4U 
1705--52 from modelling of X-ray bursts (Haberl \& Titarchuk 1994). 
This gives the $S = 1$ point at 5, 4--10 and 10--17 per cent of 
Eddington luminosity ($L_{\rm Edd}=1.76\times 
10^{38}$~ergs~s$^{-1}$), for 4U~1608--52, Aql X-1 and 4U~1705-44, 
respectively. Since luminosity estimates from X-ray bursts are 
approximate and might contain unknown systematic errors (see e.g.\ in 
't Zand et al.\ 2001), it is possible that the position in the 
colour-colour diagram of all these three atolls depend on the X-ray 
luminosity in the same way, and that the $S=1$ point corresponds to 
$\sim$ 5--10 per cent of $L_{\rm Edd}$.

\section{Discussion}

The three atolls studied here are transient systems, where the 
luminosity changes by a factor of $\ga 100$, unlike the majority of 
atolls which are not very variable. The transients go down to very 
low luminosities where they show a new track on the colour-colour 
diagram which extends the previously known atoll (or C) shaped path 
into a Z. This new upper branch is clearly distinguished from a 
simple extension of the previously known C-shaped track.  Below 
$S=1$, the track turns so that the average source luminosity 
decreases from right to left on the colour-colour diagram.  We 
propose that all atolls would show such a track if their mass 
accretion rate could change by a large enough factor, but that their 
normal, rather small range in variability limits their observed 
colour-colour diagram to only a small section of the track.

A good example of this is 4U~1728--34, regarded as an archetypal 
atoll tracing a C-shaped track in the diagram (M{\'e}ndez \& van der 
Klis 1999; Di Salvo et al.\ 2001). However, the ratio of the highest 
to lowest count rate of this source is only about 2.5 (Di Salvo et 
al.\ 2001). Therefore, we suggest that the 4U~1728--34 data collected 
so far shows a colour-colour diagram which is limited to the diagonal 
and lower branches of a Z track. The source never goes to low enough 
luminosities to sample much of the upper branch.

Thus, we show that atolls share the same colour-colour topology with 
the Z sources. With increasing accretion rate they both trace out a 
similar Z-shaped pattern. Despite this similarity, these two LMXB 
categories are of course not the same. They differ in the luminosity 
range required to cover the whole Z track (Z sources are much less 
variable then the three transient atolls presented here) and in the 
time taken to trace out the upper and diagonal branches (Z sources 
move much faster). Another important difference is the luminosity at 
which the Z-shaped pattern arises. For the Z sources the transition 
from the upper (horizontal) branch to the diagonal (normal) branch 
occurs at around 0.5--1 $L_{\rm Edd}$, while for the atolls this 
transition occurs at luminosities about ten times less. There is also 
a significant difference in the spectral shape in the upper branch: 
spectra of the atolls are much harder, similar to hard state of black 
hole candidates.

The differences in luminosity and spectral shape can be reconciled in 
a model in which the fundamental difference between atolls and Z 
sources is magnetic field. Evolution along the Z track is caused by 
the increasing mass accretion rate, $\dot{M}$, decreasing the inner 
radius, $R_{\rm in}$, of a truncated disc. For the atolls the disc is 
probably truncated by evaporation (e.g.\ R{\'o}{\.z}a{\'n}ska \& 
Czerny 2000), leading to an inner, optically thin, hot flow which 
gives the hard X-ray spectrum. However, the evaporation efficiency 
decreases as a function of increasing mass accretion rate, so this 
cannot truncate the disc in the Z sources. Instead the truncation is 
likely to be caused by stronger magnetic field, but here the 
increased mass accretion rate means that the inner flow is much more 
optically thick, and so cooler.

We note that after this paper was submitted to MNRAS, another group 
independently presented very similar results (Muno, Remillard \& 
Chakrabarty, astro-ph/0111370).

\section*{Acknowledgements}

We thank Didier Barret for attracting our attention to the 4U 
1705--44 data. We also thank the referee, Michiel van der Klis, and 
Mariano M{\'e}ndez for helpful discussions. This research has been 
supported in part by the Polish KBN grant 2P03D00514.


\label{lastpage}


\begin{thebibliography}{}



\bibitem[]{cbc01} Church M. J., Ba{\l}uci{\'n}ska-Church M., 2001, A\&A, 369, 
915

\bibitem[]{cui98} Cui W., Barret D., Zhang S. N., Chen W., Boirin L., Swank J., 
1998, ApJ, 502, L49

\bibitem[]{dis01} Di Salvo T., M{\'e}ndez M., van der Klis M., Ford 
E., Robba N. R., 2001, ApJ, 546, 1107

\bibitem[]{fk01} Fender R. P., Kuulkers E., 2001, MNRAS, 324, 923

\bibitem[]{fkk98} Ford E. C., van der Klis M., Kaaret P., 1998, ApJ, 
498, L41

\bibitem[]{for99}  Ford E. C., van der Klis M., M{\'e}ndez M., van Paradijs J., 
Kaaret P., 1999, ApJ, 512, L31 

\bibitem[]{flm89} Fortner B., Lamb F. K., Miller G. S., 1989, Nature, 342, 775

\bibitem[]{ht94} Haberl F., Titarchuk L., 1994, A\&A, 299, 414

\bibitem[]{hk89} Hasinger G., van der Klis M., 1989, A\&A, 225, 79

\bibitem[]{hom01} Homan J., van der Klis M., Jonker P. G., Wijnands
R., Kuulkers E., M{\'e}ndez M., Lewin W. H. G., 2001, astro-ph/0104323

\bibitem[]{lam89} Lamb F. K., 1989, in Proc. 23 ESLAB Symp. on X-Ray
Astronomy, ed. N. E. White, ESA SP-296 (Noordwijk: ESA)

\bibitem[]{zan01} in 't Zand J. J. M., et al., 2001, astro-ph/0104285

\bibitem[]{lan87} Langmeier A., Sztajno M., Hasinger G., Tr{\"u}mper J., 
Gottwald M., 1987, ApJ, 323, 288

\bibitem[]{men99} M{\'e}ndez M., 1999, in Proceedings of the 19$^{\rm th}$ Texas 
Symposium in Paris, published in electronic form

\bibitem[]{mk99} M{\'e}ndez M., van der Klis M., 1999, ApJ, 517, L51



\bibitem[]{mkf99} M{\'e}ndez M., van der Klis M., Ford E. C., Wijnands R., van 
Paradijs J., 1999, ApJ, 511, L49




\bibitem[]{nak89} Nakamura N., Dotani T., Inoue H., Mitsuda K., Tanaka Y., 
Matsuoka M.,1989, PASJ, 41, 617

\bibitem[]{n00} Nowak M. A., 2000, MNRAS, 318, 361

\bibitem[]{pen89} Penninx W., Damen E., Tan J., Lewin W. H. G., van Paradijs J., 
1989, A\&A, 208, 146

\bibitem[]{ps95} Predehl P., Schmitt J. H. M. M., 1995, A\&A, 293, 889

\bibitem[]{ps01} Psaltis D., Norman C., 2000, astro-ph/0001391

\bibitem[]{rei00} Reig P., M{\'e}ndez M., van der Klis M., Ford E. 
C., 2000, ApJ, 530, 916

\bibitem[]{rc00} R{\'o}{\.z}a{\'n}ska A., Czerny B., 2000, MNRAS, 316, 473

\bibitem[]{rut01} Rutledge R. E., Bildsten L., Brown E. F., Pavlov G. 
G., Zavlin V. E., 2001, ApJ, 559, 1054

\bibitem[]{kli95} van der Klis M., 1995, in Lewin W. H.  G., van Paradijs J., 
van den Heuvel E. P. J., eds., X-ray Binaries, Cambridge University Press, 
Cambridge, P.\ 252

\bibitem[]{kli00} van der Klis M., 2000, ARAA, 38, 717

\bibitem[]{kli01} van der Klis M., 2001, ApJ, 561, 943

\bibitem[]{wk99} Wijnands R., van der Klis M., 1999, ApJ, 514, 939 





\end{thebibliography}
\end{document}